\documentclass[%
 aip,
 apl,%
 amsmath,amssymb,
 reprint,%
]{revtex4-1}
\usepackage{mhchem}
\usepackage{graphicx}
\usepackage{dcolumn}
\usepackage{bm}

\begin{document}
\preprint{AIP/123-QED}

\title{Band dependence of charge density wave in quasi-one-dimensional \ce{Ta_2NiSe_7} probed by orbital magnetoresistance}%

\author{Jiaming He}
\affiliation{Key Laboratory of Artificial Structures and Quantum Control and Shanghai Center for Complex Physics,
School of Physics and Astronomy, Shanghai Jiao Tong University, Shanghai $\text{200240, China}$}

\author{Yiran Zhang}
\affiliation{Key Laboratory of Artificial Structures and Quantum Control and Shanghai Center for Complex Physics,
School of Physics and Astronomy, Shanghai Jiao Tong University, Shanghai $\text{200240, China}$}

\author{Libin Wen}
\affiliation{Key Laboratory of Artificial Structures and Quantum Control and Shanghai Center for Complex Physics,
School of Physics and Astronomy, Shanghai Jiao Tong University, Shanghai $\text{200240, China}$}

\author{Yusen Yang}
\affiliation{Key Laboratory of Artificial Structures and Quantum Control and Shanghai Center for Complex Physics,
School of Physics and Astronomy, Shanghai Jiao Tong University, Shanghai $\text{200240, China}$}

\author{Jinyu Liu}
\affiliation{Department of Physics, Tulane University, New Orleans, LA 70118, USA}

\author{Yueshen Wu}
\affiliation{Key Laboratory of Artificial Structures and Quantum Control and Shanghai Center for Complex Physics,
School of Physics and Astronomy, Shanghai Jiao Tong University, Shanghai $\text{200240, China}$}

\author{Hailong Lian}
\affiliation{Key Laboratory of Artificial Structures and Quantum Control and Shanghai Center for Complex Physics,
School of Physics and Astronomy, Shanghai Jiao Tong University, Shanghai $\text{200240, China}$}

\author{Hui Xing}
\email{huixing@sjtu.edu.cn}
\affiliation{Key Laboratory of Artificial Structures and Quantum Control and Shanghai Center for Complex Physics,
School of Physics and Astronomy, Shanghai Jiao Tong University, Shanghai $\text{200240, China}$}

\author{Shun Wang}
\affiliation{Key Laboratory of Artificial Structures and Quantum Control and Shanghai Center for Complex Physics,
School of Physics and Astronomy, Shanghai Jiao Tong University, Shanghai $\text{200240, China}$}

\author{Zhiqiang Mao}
\affiliation{Department of Physics, Tulane University, New Orleans, LA 70118, USA}

\author{Ying Liu}
\email{yingl@sjtu.edu.cn}
\affiliation{Key Laboratory of Artificial Structures and Quantum Control and Shanghai Center for Complex Physics,
School of Physics and Astronomy, Shanghai Jiao Tong University, Shanghai $\text{200240, China}$}
\affiliation{Collaborative Innovation Center of Advanced Microstructures, Nanjing $\text{210093, China}$}
\affiliation{Department of Physics and Materials Research Institute, Pennsylvania State University, University Park, Pennsylvania 16802, USA}

\begin{abstract}
\ce{Ta2NiSe7} is a quasi-one-dimensional (quasi-1D) transition-metal chalcogenide with
 Ta and Ni chain structure. An incommensurate charge-density wave (CDW) in this quasi-1D structure was well studied previously using tunnelling spectrum, X-ray and electron diffraction, whereas its transport property and the relation to the underlying electronic states remain to be explored. Here we report our results of magnetoresistance (MR) on \ce{Ta2NiSe7}. A breakdown of the Kohler's rule is found upon entering the CDW state. Concomitantly, a clear change of curvature in the field dependence of MR is observed. We show that the curvature change is well described by two-band orbital MR, with the hole density being
 strongly suppressed in the CDW state, indicating that the $p$ orbitals from Se atoms dominate the
 change in transport through the CDW transition.
\end{abstract}
\maketitle

Low-dimensional transition-metal chalcogenides (TMC) garnered great interests due to their rich
physical properties \cite{Wang2012, Wang2015}, including the recent discovery of valley dependent
transport \cite{Mak2012} and superconductivity \cite{Lu2015} in MoS$_2$, extremely large
MR in WTe$_2$ \cite{Ali2014} and the stunning topological phases \cite{Qi2011}. The low
dimensionality and high structural symmetry makes TMC ideal for studying the structure-property relation.
Uniquely, the electron-electron interaction in 1D metals causes strong perturbations, in sharp contrast to the case in the 2D and 3D Fermi-liquid counter parts, and leads to Luttinger liquid (LL) behavior \cite{Luttinger1963}. On the other hand, CDW in the quasi-1D metal can be viewed as a classical analogue of a LL state \cite{Slot2004}. These facts raise special interests in the study of CDW in 1D metallic systems for exploring emergent behaviours. In addition, strong nonlinear electrical transport due to CDW sliding in 1D chain systems leads to high dielectric constants \cite{Zant1996} and narrow-band noise \cite{Zybtsev2009}, both of which may find interesting use in applications.

\ce{Ta2NiSe7} is a quasi-1D ternary TMC showing an incommensurate CDW \cite{Fleming1990}. Its quasi-1D structure is illustrated in Fig. 1(a).  Similar to \ce{FeNb_3Se_10} \cite{Meerschaut1981}, the unit cell of \ce{Ta2NiSe7} consists of double rows of tantalum atoms (Ta1) in bicapped trigonal prismatic selenium coordination, and the other double rows of
tantalum atoms (Ta2) in octahedral selenium coordination; nickel atoms are in highly distorted
octahedral coordination. Band structure calculations showed that the Fermi surface consists of contribution from Ta2
$d$ orbitals in the octahedral chains and Se2 $p$ orbitals from trigonal prismatic columns \cite{Canadell1987}.
At around 52 K, an incommensurate CDW occurs. Interestingly, the CDW formation mechanism was suggested
not to be the Fermi surface nesting effect, but rather, through the charge transfer between the two Ta2 chains \cite{STM1992}.
However, X-ray diffraction experiment pointed out that
all Ta2 atoms are equivalent in symmetry in the CDW state\cite{Xray1995}. So far, this discrepancy is not fully resolved, but is mitigated by the observation of two independent CDWs, with modulation wave vector $2k_\text{F}$ and $4k_\text{F}$, each corresponding to the transverse displacement of Ni and Se2 \cite{Xray1995}, and the longitudinal modulation of Ta2 \cite{Ludecke2000}, respectively.

The CDW in \ce{Ta2NiSe7} is found to be sensitive to defects \cite{Fleming1990, Ludecke2000Nb}. The typical role of defects on CDW includes scattering carriers, pinning the CDW domains and jeopardising the formation of
long-range structural order. In 1D systems, the situation can be more complicated. The spacial distribution of defects can be driven by the CDW order to show a spacial distribution with the same modulation wave vector, which is the so-called defect quasiregularity induced by CDW \cite{Baldea1993}. The consequence, such as the thermal hysteresis of the CDW transition \cite{Baldea1990}, was indeed found in \ce{Ta2NiSe7} \cite{Fleming1990}.
CDW gap only removes part of the Fermi surface, so that the system remains semimetallic below CDW transition temperature ($T_\text{CDW}$) \cite{Canadell1987, STM1992}.

Regarding the transport property of \ce{Ta2NiSe7},  so far only a small kink in the
temperature dependence of resistivity $R(T)$ was found to correspond to the CDW \cite{Fleming1990}. The transport related to the 1D character, and the underlying electronic states are not explored.
Here, we report the result of MR measurement on high-quality \ce{Ta2NiSe7} single crystals.
In addition to a sizable MR up to $35\%$ at low temperatures, we find that the Kohler's scaling is valid at high temperatures but fails in the CDW state. A clear change of curvature in the field dependence of MR upon entering the CDW state is also observed. The behavior is well interpreted using a two-band orbital MR model,
which shows that the hole density is strongly suppressed in the CDW state, while the electron density is
less affected. Our results provide transport evidences showing that CDW occurs mostly in hole-like
band from Se $p$ orbitals.

\ce{Ta2NiSe7} single crystals were prepared by flux method. X-ray diffraction (XRD) were performed in
Bruker D8 diffractometer. XRD on selected samples showed a space
group of $(\text{}C2/m)$, and lattice constants of $a = 13.84 \  \AA$, $b = 3.48 \  \AA$, $c =
18.60 \  \AA$ , $\alpha = \gamma = 90^{\circ}$ , $\beta = 108.8^{\circ}$, consistent with earlier
report \cite{Sunshine1986}. Single crystal morphology and elemental analysis were carried out by
scanning electron microscope and energy dispersive X-ray spectroscopy. Crystals used in our measurement are highly selected, with a residual resistance ratio $RRR > 7$, which is the highest among those reported in literature. Typical size of the crystal is \(200 \times 50\times20\ \mu m^3\).
$R_b$, the resistance along the $b$ axis, was measured by standard four terminal method in a Quantum
Design PPMS system with a 14 Tesla magnet, with a rotator for controlling the relative orientation between
magnetic field and crystal.

\begin{figure} \includegraphics[width=1.0\columnwidth]{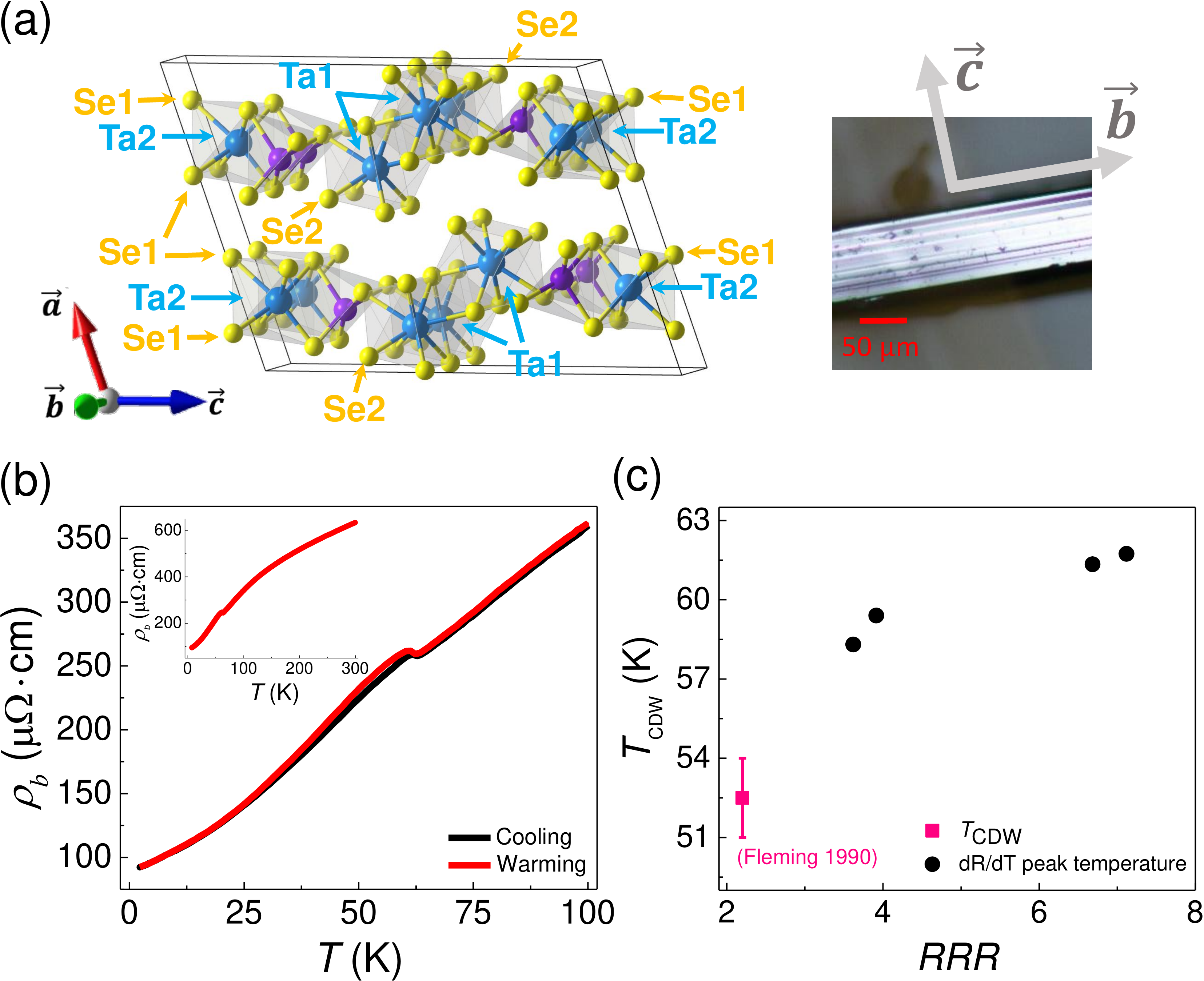} \caption{(a) Left: a schematic of \ce{Ta2NiSe7} unit cell. Right: an optical image of a \ce{Ta2NiSe7} crystal showing a naturally cleaved $bc$ surface. (b) Temperature dependence of resistivity from 2 to 100 K, red (black) curve is measured during warming (cooling). Inset shows the temperature dependence of resistivity from 2 to 300 K. (c) $RRR$ versus the CDW transition temperature $T_\text{CDW}$. $T_\text{CDW}$ is determined using the peak temperature of $dR/dT$ curve.} \end{figure}

\ce{Ta2NiSe7} shows a metallic behavior in the temperature range of 2 to 300 K, as shown in Fig. 1(b).
$T_\text{CDW}$, defined by the peak temperature in d$R$/d$T$,
is at around 62.5 K. This value is nearly 10 K higher
than the 52.5 K reported in earlier literatures \cite{Fleming1990, STM1992, Xray1995} and comparable to the highest reported so far \cite{Ludecke2000Nb}. The large difference in $T_\text{CDW}$ is attributed to different sample qualities due to the existence of impurities. Typically, this can be characterised using $RRR$. We studied several samples grown in different batches and obtained the relation between $T_\text{CDW}$ and $RRR$ in Fig. 1(c). The value from an earlier study \cite{Fleming1990} is also included in the plot. It is clear to see that $T_\text{CDW}$ increase monotonically with increasing $RRR$. In this paper, we focus on the sample with the highest $T_\text{CDW}$ of 62.5 K, which corresponds to the lowest impurity concentration. Another interesting feature in $R(T)$ is the hysteresis loop below $T_\text{CDW}$.  One should note that the resistivity upon warming is higher than that upon cooling. This feature was attributed to the defect quasi regularity induced by CDW, as described above.

\begin{figure} \includegraphics[width=1.0\columnwidth]{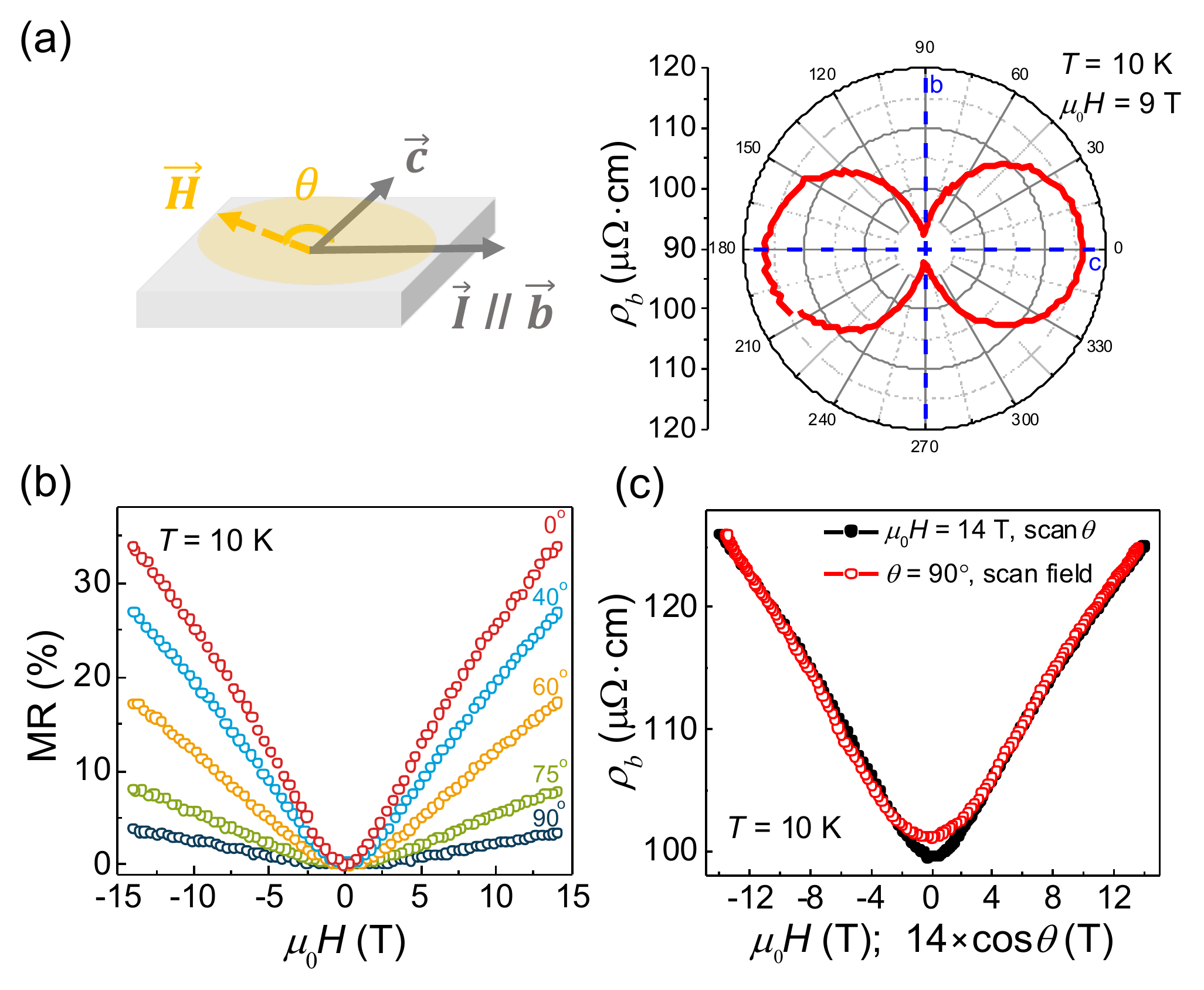} \caption{(a) Left: a schematic of measurement configuration. The magnetic field is rotating in the $bc$ plane, $\theta$ was defined by the angle between magnetic field and $c$ axis. Right: polar plot of MR vs. $\theta$ in a 9 T field rotating in $bc$ plane. (b) Field dependence of MR at 10 K with field applied along different directions in the $bc$ plane. (c) Black dots: the field dependence of $\rho_b(H)$; Red circles: $\rho_b$ vs. $14 \times \text{cos}\theta$ which is the perpendicular component of a 14 T field. Both are measured at 10 K.} \end{figure}

\begin{figure} \includegraphics[width=1.0\columnwidth]{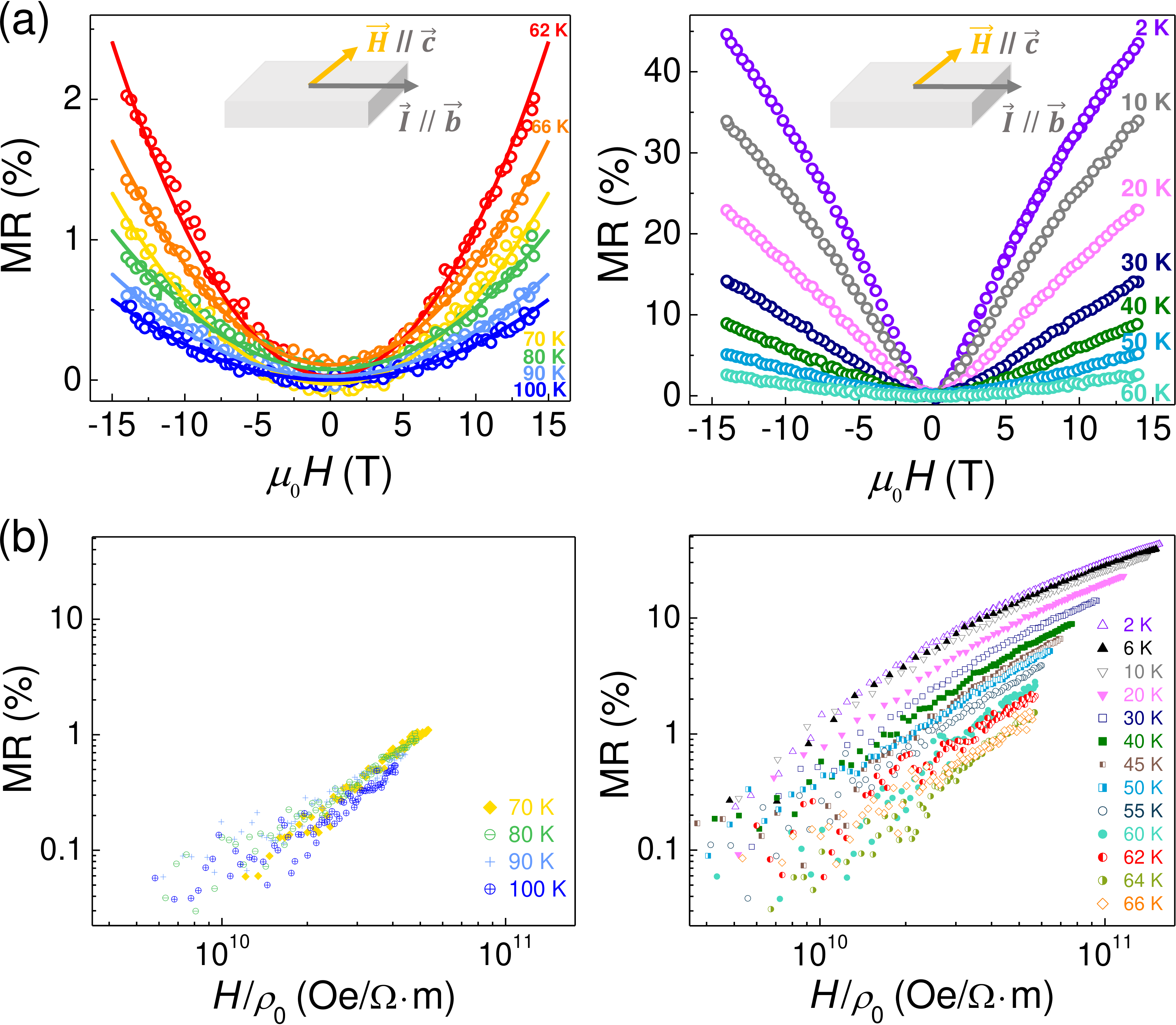} \caption{(a) Field dependence of MR measured along the $b$ axis in a magnetic field along the $c$ axis at different temperatures. Solid lines in left panel are fittings using parabolic field dependence. (b) The Kohler's plot for temperatures above and below the CDW transition temperature.} \label{}
\end{figure}

Now we focus on the MR of \ce{Ta2NiSe7}. Shown in Fig. 2 are the MR measured along the $b$ axis under a rotating magnetic field in the $bc$ plane. The polar plot of $R(\theta)$ in Fig. 2(a), measured at 10 K under a 9 T field, shows a pronounced two-fold symmetry, with symmetry axis along $b$/$c$ direction. In this configuration, different part of the Fermi surface is probed when the magnetic field rotates along different directions. To explore the nature of the MR at low temperatures, the field dependence of MR at 10 K is shown in Fig. 2(b). MR is largest with field perpendicular to the current ($\theta = 0$), reaching 35\% at 14 T, and decreases progressively to below 4\% when the field is parallel to the current. Clearly the orbital MR (also know as the ordinary MR) is dominating. We further illustrate this point in Fig. 2(c) by comparing two curves: $\rho_b(H)$ for field along $c$ axis and $\rho_b$ $vs.$ $14 \times \text{cos}(\theta)$ under a rotating field of 14 T. For the latter,  $14 \times \text{cos}(\theta)$ is essentially the field component perpendicular to the current direction. The two
curves overlap rather well, meaning that the field perpendicular to current is critical, which
further proves that the effect is dominated by orbital MR. For large $\theta$ region in $\rho_b$ $vs.$ $14\times\text{cos}(\theta)$ curve, the field is oriented close to the $b$ axis, which in turn probe a different area of Fermi surface, as shown in Fig. 2(a). This leads naturally to the difference between the two curves in the large
$\theta$ region (small field region in $\rho_b(H)$ curve).

In an in-plane magnetic field perpendicular to current, the field dependence of MR at different temperatures is shown in Fig. 3(a). One major observation is that MR$(H)$ shows concave and convex curvature for temperatures above and below $T_\text{CDW}$, respectively. The slope change in the field dependence of ordinary MR signifies a multiband effect, consist with the semimetallic band structure found in earlier calculation \cite{Canadell1987}. At high temperatures MR features a parabolic field dependence, again indicating that orbital MR is dominating, a mechanism the same with that at low temperatures.

We find that MR data above $T_\text{CDW}$ follows the Kohler's rule \cite{Ziman1972}, namely, $\Delta \rho/\rho_0 = f(H/\rho_0)$. Here $\rho_0$ is the zero-field resistivity at a given temperature, and $f(H/\rho_0)$ stands for a function of $H/\rho_0$. The Kohler's rule typically holds well for materials with carrier density insensitive to temperature and with isotropic scattering. In Fig. 3(b), data in \ce{Ta2NiSe7} for $T > T_\text{CDW}$ falls on a master curve of Kohler scaling, with an exponent of 1.99, as expected for an orbital MR. However, when the temperature is lowered through $T_\text{CDW}$, significant deviation develops, leading to the scattered curves in the Kohler's plot, as shown in the right panel of Fig. 3(b). In this multi-band system, the breakdown of Kohler's rule in the CDW states indicate that either the ratio of electron and hole carrier changes strongly, or the scattering becomes anisotropic. While both possibilities may be relevant, the former appears to be more tempting since a direct consequence of a CDW order is the removal of carriers.

\begin{figure} \includegraphics[width=1.0\columnwidth]{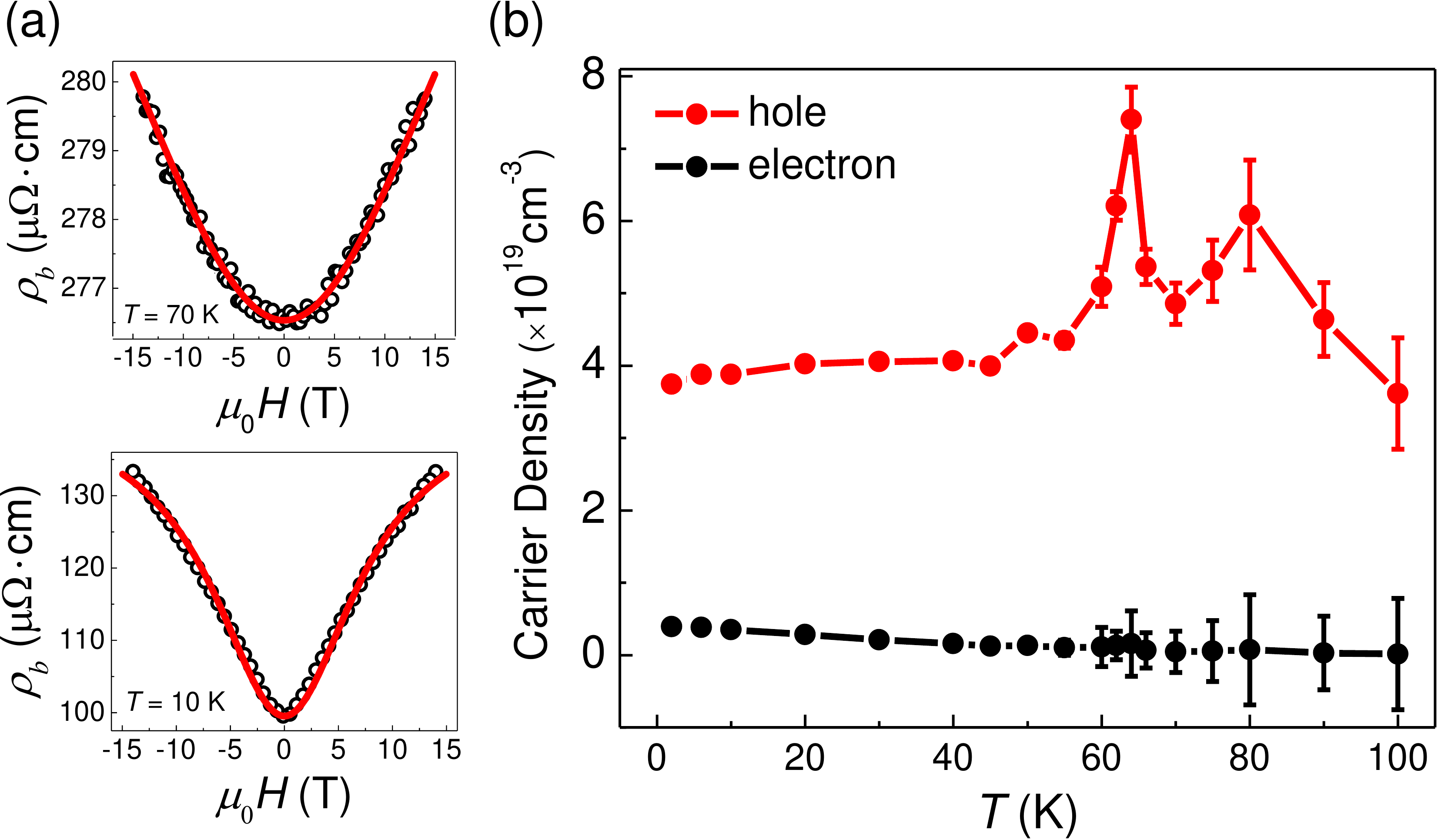} \caption{(a) Field dependence of MR at 10 and 70 K, showing different �curvature. Red lines are fitting curves using the two-band resistivity model. (b) Hole and electron density as a function of temperature, obtained from the fitting (see main text).}
\label{} \end{figure}

Here we attempt to provide a more quantitative understanding of the MR using the semiclassical
two-band resistivity,
\[\rho_{xx} =
\text{Re}(\rho)=\frac{1}{e}\frac{(n_h\mu _h+n_e \mu _e) + (n_h\mu _e + n_e\mu _h)\mu _h\mu_eB^2}{(n_h \mu
_h + n_e \mu _e)^2 + (n_h-n_e)^2\mu _h^2 \mu _e^2B^2} \]%
Here, $e$ is the electron charge, $n_e$
and $n_h$ are the  electron and hole density, $\mu_e$ and $\mu _h$ are the electron and hole mobility, respectively.
With this equation, we can fit the MR at different temperatures and extract the transport quantities. Unfortunately, because of the inclusion of four parameters, it is difficult to get a reliable fitting if we set all the four parameters independent. Now with the implication of the above discussions, especially the clue we obtained in the Kohler scaling, we intentionally set the mobility $\mu$ of electron and hole to be identical to facilitate an easier fitting. Fig. 4(a) shows the representative MR data and the fitting curve. One can see that the two-band ordinary MR equation describes the data quite well, for temperatures both above and below $T_\text{CDW}$. The resultant fitting parameters are summarized in Fig. 4(b).
One will also note that electron and hole are entirely symmetric in the two-band resistivity model, therefore it is
impossible to differentiate the two solely from the model. For this we rely on the Hall
resistivity data of \ce{Ta2NiSe7}, and found that the hole dominates over the entire temperature range
(data not shown here), which allows us to distinguish $n_e$ and $n_h$ in Fig. 4(b). Consistently the resultant hole density is several times higher than the electron density over the entire temperature range. We see that the carrier density shows clearly different behavior for electron-like and hole-like bands: the electron density increases only slightly upon cooling, while the hole density decreases
significantly when entering the CDW state.  This contrasting behavior shows unambiguously that the CDW gap open predominately in the hole-like band. Recall that the Fermi surface of \ce{Ta2NiSe7} consists of an electron-like band from Ta2 $d$ orbitals in the octahedral chains, and a hole-like band Se2 $p$ orbitals from trigonal prismatic columns \cite{Canadell1987},
our results indicate that the carrier transport is dominated by hole-like band from the $p$ orbitals from Se2 atoms, which finds good consistency with earlier synchrotron X-ray diffraction result that the primary part of CDW resides on Se2 and Ni \cite{Xray1995}.

In summary, we report our magneto resistivity measurement on the quasi-1D  transition-metal chalcogenide \ce{Ta2NiSe7}. The Kohler's rule is valid at high temperatures, but breaks down in the CDW state. A clear change
of curvature in the field dependence of magnetoresistivity upon entering the CDW state is observed, which is
fully accounted for by a two-band ordinary magnetoresistivity model. The two-band fitting shows
that the hole carrier density is strongly suppressed in the CDW state, indicating that CDW takes place in
hole-like band, mostly the $p$ orbitals from Se2 atoms.

The work at SJTU is supported by MOST (Grant No. 2015CB921104) and NSFC (Grant Nos. 91421304 and 11474198), at Penn State by NSF (Grant No. EFMA1433378), at Tulane supported by the U.S. Department of Energy under EPSCoR Grant No. DE-SC0012432 with additional support from the Louisiana Board of Regents (support for a graduate student, materials, travel to NHMFL).


\begin{thebibliography}{21}
\bibitem{Wang2012} Q. H. Wang, K. Kalantar-Zadeh, A. Kis, J. N. Coleman, and M. S. Strano, ``Electronics and optoelectronics of two-dimensional transition metal dichalcogenides,"\ \href{http://dx.doi.org/10.1038/nnano.2012.193}{Nat Nano \textbf{7}, 699–712 (2012).}
\bibitem{Wang2015} H. Wang, H. Yuan, S. Sae Hong, Y. Li, and Y. Cui, ``Physical and chemical tuning of two-dimensional transition metal dichalco-genides,"\ \href{http://dx.doi.org/10.1039/C4CS00287C}{Chem. Soc. Rev. \textbf{44}, 2664–2680 (2015).}
\bibitem{Mak2012} K. F. Mak, K. He, J. Shan, and T. F. Heinz, ``Control of valley polarization in monolayer MoS2 by optical helicity,"\ \href{http://dx.doi.org/10.1038/nnano.2012.96}{Nat Nano \textbf{7}, 494–498 (2012).}
\bibitem{Lu2015} J. M. Lu, O. Zheliuk, I. Leermakers, N. F. Q. Yuan, U. Zeitler, K. T. Law, and J. T. Ye, ``Evidence for two-dimensional Ising superconductivity in gated MoS2,"\ \href{http://dx.doi.org/10.1126/science.aab2277%7D}{Science \textbf{350}, 1353–1357 (2015).}
\bibitem{Ali2014} M. N. Ali, J. Xiong, S. Flynn, J. Tao, Q. D. Gibson, L. M. Schoop, T. Liang, N. Haldolaarachchige, M. Hirschberger, N. P. Ong, and R. J. Cava, ``Large, non-saturating magnetoresistance in WTe2,"\ \href{http://dx.doi.org/10.1038/nature13763%7D}{Nature \textbf{514}, 205 (2014).}
\bibitem{Qi2011} X.-L. Qi and S.-C. Zhang, ``Topological insulators and superconductors,"\ \href{http://dx.doi.org/10.1103/RevModPhys.83.1057}{Rev. Mod. Phys. \textbf{83}, 1057–1110 (2011).}
\bibitem{Luttinger1963} J. M. Luttinger, ``An Exactly Soluble Model of a Many‐Fermion System,"\ \href{http://dx.doi.org/10.1063/1.1704046}{Journal of Mathematical Physics \textbf{4}, 1154–1162 (1963).}
\bibitem{Slot2004} E. Slot, M. A. Holst, H. S. J. van der Zant, and S. V. Zaitsev-Zotov,``One-Dimensional Conduction in Charge-Density-Wave Nanowires,"\href{http://dx.doi.org/10.1103/PhysRevLett.93.176602}{Phys. Rev. Lett. \textbf{93}, 176602 (2004).}
\bibitem{Zant1996} H.\ S.\ J. van der Zant and O. C. Mantel and C. Dekker and J. E. Mooij and C. Tr{\ae}holt,``Thin‐film growth of the charge‐density‐wave oxide Rb0.30MoO3,"\href{http://dx.doi.org/10.1063/1.116629}{Applied Physics Letters \textbf{68}, 3823–3825 (1996).}
\bibitem{Zybtsev2009} S. G. Zybtsev, V. Y. Pokrovskii, V. F. Nasretdinova, and S. V. Zaitsev-Zotov,``Gigahertz-range synchronization at room temperature and other features of charge-density wave transport in the quasi-one-dimensional conductor NbS3,"\href{http://dx.doi.org/10.1063/1.3111439}{Applied Physics Letters \textbf{94}, 152112 (2009).}
\bibitem{Fleming1990} R. M. Fleming, S. A. Sunshine, C. H. Chen, L. F. Schneemeyer,and J. V.  Waszczak,``Defect-inhibited incommensurate distortion in Ta2NiSe7,"\href{}{Physical Review B \textbf{42}, 4954 (1990).}
\bibitem{Meerschaut1981} A. Meerschaut, P. Gressier, L. Guemas, and J. Rouxel,``Structural determination of a new one-dimensional niobium chalcogenide: FeNb3Se10,"\href{}{Materials Research Bulletin \textbf{16}, 1035–1040 (1981).}
\bibitem{Canadell1987} E. Canadell and M. H. Whangbo,``Metallic versus nonmetallic properties of ternary chalcogenides: tantalum metal selenide, Ta2MSe7 (M= nickel, platinum), and tantalum nickel chalcogenide, Ta2NiX5 (X= sulfide, selenide),"\href{}{Inorganic Chemistry \textbf{26}, 3974–3976 (1987).}
\bibitem{STM1992} Z. Dai, C. G. Slough, W. W. McNairy, and R. V. Coleman,``Charge-density-wave formation in Ta2NiSe7 studied by scanning tunneling microscopy,"\href{}{Physical review letters \textbf{69}, 1769–1772 (1992).}
\bibitem{Xray1995} A. Spijkerman, A. Meetsma, J. L. de Boer, Y. Gao, and S. van Smaalen,``Modulated structure of NiTa 2 Se 7 in its incommensurate charge-density-wave state at 16 K,"\href{}{Physical Review B \textbf{52}, 3892 (1995).}
\bibitem{Ludecke2000} J. L{\"u}decke, M. Schneider, and S. van Smaalen,``Independent q→ and 2q→ Distortions in the Incommensurately Modulated Low-Temperature Structure of NiTa2Se7,"\href{http://dx.doi.org/10.1006/jssc.2000.8766}{Journal of Solid State Chemistry \textbf{153}, 152 – 157 (2000).}
\bibitem{Ludecke2000Nb} J. L{\"u}decke, E. Riedl, M. Dierl, K. Hosseini, and S. van Smaalen,``Influence of niobium on the charge-density-wave transition of $\mathrm{Ni}({\mathrm{Ta}}_{1\ensuremath{-}x}{\mathrm{Nb}}_{x}{)}_{2}{\mathrm{Se}}_{7}$,"\href{http://dx.doi.org/10.1103/PhysRevB.62.7057}{Phys. Rev. B \textbf{62}, 7057–7062 (2000).}
\bibitem{Baldea1993} I. B\^aldea and M. B\ifmmode \u{a}\else \u{a}\fi{}descu,``Quasiregular impurity distribution driven by a charge-density wave,"\href{http://dx.doi.org/10.1103/PhysRevB.48.8619}{Phys. Rev. B \textbf{48}, 8619–8628 (1993).}
\bibitem{Baldea1990} I. B\^aldea,``Reentrant charge-density wave in the one-dimensional system with impurities - a self-consistent approach,"\href{}{Physica Scripta \textbf{42}, 749 (1990).}
\bibitem{Sunshine1986} S. A. Sunshine and J. A. Ibers,``Synthesis, structure, and transport properties of tantalum nickel selenide (Ta2NiSe7) and tantalum platinum selenide (Ta2PtSe7),"\href{}{Inorganic Chemistry \textbf{25}, 4355–4358 (1986).}
\bibitem{Ziman1972} J. Ziman,\textit{Principles of the Theory of Solids} \href{}{(Cambridge University Press, 1972).}

\end{thebibliography}
\end{document}